\documentclass{iaus}
\usepackage{graphics}
\usepackage[dvips]{graphicx}

  \checkfont{eurm10}
  \iffontfound
    \IfFileExists{upmath.sty}
      {\typeout{^^JFound AMS Euler Roman fonts on the system,
                   using the 'upmath' package.^^J}%
       \usepackage{upmath}}
      {\typeout{^^JFound AMS Euler Roman fonts on the system, but you
                   dont seem to have the}%
       \typeout{'upmath' package installed. iaus.cls can take advantage
                 of these fonts,^^Jif you use 'upmath' package.^^J}%
      }
  \else
  \fi

  \checkfont{msam10}
  \iffontfound
    \IfFileExists{amssymb.sty}
      {\typeout{^^JFound AMS Symbol fonts on the system, using the
                'amssymb' package.^^J}%
       \usepackage{amssymb}%

      }{}
  \fi

  \IfFileExists{amsbsy.sty}
    {\typeout{^^JFound the 'amsbsy' package on the system, using it.^^J}%
     \usepackage{amsbsy}}
    {}

\newsavebox{\astrutbox}
\sbox{\astrutbox}{\rule[-5pt]{0pt}{20pt}}

\title[A Computer Science approach]{Computer Science approach to the stellar 
fabric of violent starforming regions in AGN}

\author[E. Terlevich {\it et al.\/}]%
{Elena Terlevich$^1$, R. Terlevich$^1$,
J. P. Torres Papaqui$^1$,
T. Estrada Piedra$^1$,
O. Fuentes $^1$,
T. Solorio$^1$,
\and S. Bressan $^{2,1}$
 }

\affiliation{$^1$Instituto Nacional de Astrof\'\i sica, Optica y 
Electr\'onica, 72840 Puebla, MEXICO \\
 email: eterlevi,rjt,fuentes,papaqui,thamar,trilce@inaoep.mx\\[\affilskip]
$^2$ Osservatorio Astronomico di Padova, Vicolo dell'Osservatorio 535122, 
Padua, ITALY email:bressan@pd.astro.it}

\pubyear{2004}
\volume{222}
\pagerange{1--8}
\date{?? and in revised form ??}
\jname{The Interplay among Black Holes, Stars and ISM \\in Galactic Nuclei}
\editors{Th. Storchi Bergmann, L.C. Ho \& H.R. Schmitt, eds.}
\begin{document}

\maketitle

\begin{abstract}

In order to analyse  the  large numbers of Seyfert galaxy spectra available 
at present, we are testing new techniques to derive their physical parameters
fastly and accurately.

We present an experiment on such a new technique to segregate old and young  stellar populations in galactic spectra 
using machine learning methods. We used an ensemble of classifiers, each classifier in the ensemble 
specializes in young or old populations and was  trained with locally weighted regression and tested 
using ten-fold cross-validation. Since the relevant information concentrates in certain regions of 
the spectra we used the method of sequential floating backward selection offline for feature selection.

Very interestingly, the application to Seyfert galaxies proved that this technique is very insensitive to the dilution 
by the Active Galactic Nucleus (AGN) continuum. Comparing with exhaustive search we concluded that 
both methods are similar in terms of accuracy but the machine learning method is faster by about 
two orders of magnitude.
\end{abstract}

\firstsection 
\section{Introduction}

In the last decade we have witnessed a world wide boom in projects that provide clues
to elaborate a picture to describe the evolution of the Universe. They are mostly 
large observational projects that generate  huge amounts of data (2dF, 6dF, SDSS, 2MASS, VIRMOS, DEEP2... 
INAOE/UMASS will also contribute 
after the expected first light of the Large Millimeter Telescope LMT/GTM in the near future).
We are confronted with a new situation in Astrophysics for which, as a comunity, we 
are inadequately prepared.

The large number of new facilities coupled to technological advances both in
sensors and storage and in automation of data acquisition, plus a new generation of telescopes
covering from mm to X-Rays and also having multiplexing capabilities, create an unprecedented challenge:
the need to unify data, obtained both from ground and from space forming MEGADATASETS,
which lack standardization and, even worse, each database uses different access software.

There is a revolution in the way we do Astrophysics in that we need to search for and develop applications 
to automatically (and inteligently) manage, query, visualize, analyse, etc.\ the whole space and variety 
of larga datasets (including artificial or synthetic ones). This is particularly relevant for places like Mexico, 
with restricted access to large observing facilities.

Recent spectroscopic surveys of nearby AGN have proven that a large fraction show high-order hydrogen 
Balmer absorption lines in the near-UV. These features are characteristic of young stars and therefore 
represent strong evidence of recent star formation in the nuclear regions of these galaxies. 

From a theoretical point of view, it is very important to determine the age of these starbursts, 
in order to understand the nature of the starburst-AGN connection and galaxy formation and evolution. 
The characterization of the nuclear star forming region (its age and mass) is very difficult to achieve 
in AGN, due to the contamination of the nuclear stellar absorption lines by the AGN component itself. 
The recent release of high-resolution spectra of large number of galaxies by the Sloan Digital 
Sky Survey (SDSS) consortium allows spectroscopic studies to be performed now on thousands of 
galaxies with  active nuclei.

As members of a European/Mexican `Violent star formattion' network, we have started a project to determine the 
composition of the stellar population of galaxies from their spectra (\cite[Solorio et al 2004]{Thamy04}) 
and in particular, 
the nuclear stellar population of AGN using machine learning methods, that will allow us
to segregate old and young population in spectra, and apply the method to large number of 
objects (e.g.\ SDSS) at no prohibitive computational cost.

\section{The method}

The method applied here  creates and uses an ensemble of classifiers (each one
specializes in ``young'' or ``old'' stellar populations) to train the system via 
Locally Weighted Regression, which is much faster than neural-networks).
The fact that relevant information concentrates in certain regions of
the spectra (high order Balmer absorptions for the young and CaII K lines for the old populations)
helps to chose a fast method for future selection.
The algorithms are briefly described in~\S\,\ref{sfs}, \ref{lwr} and \ref{ensembles}.

\subsection{Sequential Floating Backward Selection (SFBS)} \label{sfs}

This is a feature selection algorithm that allows to work with non-monotonic data. 
It constructs in parallel the feature sets of all dimensionalities up to a specified
 threshold and consists of applying after each feature exclusion a number of features 
inclusion as long as the resulting subsets are better than those previously evaluated 
at that level. It makes a dynamically controlled number of iterations and achieves good 
results without static parameters (\cite[Pudil 1994]{Pudil94}).

\subsection{Locally Weighted Regression (LWR)} \label{lwr}

This is an instance based learning method; it assumes that instances can be represented 
as points in an Euclidean space (\cite[Schneider \& Moore 2000]{Schneider00}). Its training consists of 
explicitly retaining the training data and using them each time a prediction needs to be made. 
LWR performs a regression around a point of interest using only a local region around that point. 
Locally weighted regression can fit complex functions in an accurate way and data modifications have 
little impact on the training.

\subsection{Ensembles of Classifiers}\label{ensembles}

An ensemble of classifiers is a group of classifiers trained independently whose 
outputs are combined in some way, usually by voting (\cite[Mitchell 1997]{Mitchell97}). 
They are normally more accurate than the individual classifiers that make it up.

Several algorithmms tested rendered comparable results regarding precision, but the machine learning used
was faster by up to two orders of magnitude.

\section{Results and Conclusions}

The discovery of high order Balmer absorption lines in the near UV, characteristic
of young stars has been reported in many
spectra of Seyfert galaxies (\cite[Gonzalez-Delgado et al. 1999]{GD99}; \cite[Joguet et al 2001]{Benoit01};
\cite[Torres-Papaqui et al. 2004]{Papaqui04}).
The analysis of these features allows to estimate the age and mass of these nuclear 
starbursts, task that is otherwise
very difficult to achieve due to contamination by the non-stellar component.

Our first step in this investigation was to apply the machine learning method to synthetic 
data to study its efficiency and validity, following the diagram shown in Figure 1 (\cite[Estrada-Piedra et al.
2004]{Estrada04}).
The `data' consisted of fourteen high resolution synthetic spectra (Bressan, 
Bertone and Rodr\'\i guez, private communication) considering ten levels of dilution by a 
power law and with added Gaussian noise. The age calibration was performed using
young starburst ($10^7 - 10^{8.7}$yrs) and  old bulge  ($10^8 - 10^{9.6}$yrs) spectra
characterized by the Balmer lines and the CaII K line respectively.

\begin{figure}
 \includegraphics[width=18cm,height=12cm]{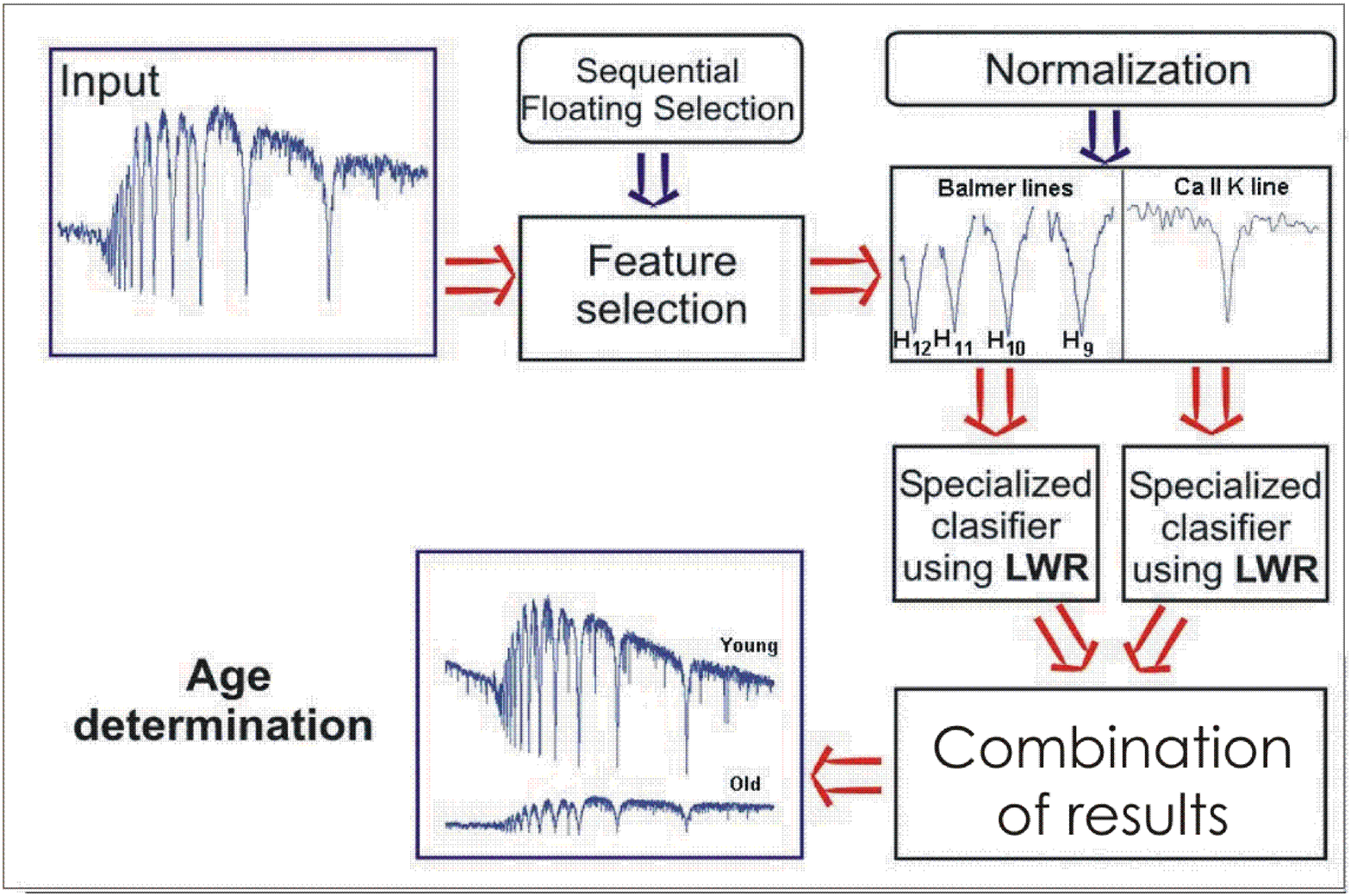}
  \caption{Schematic process of age identification.
   }
\end{figure}

The synthetic spectra, with a 0.2\AA\ sampling, covers a range of 3600 - 5300 \AA .
The accuracy achieved in recovering the input age is 0.3 dex in log(age).
To our delight, the technique proved to be quite insensitive to dilution by a featureless continuum.

The experimental results  show the efficiency of the automatic learning method applied to astronomy.
Our next step is to apply the method to 
a sample of real data and to define the training set with a young, intermediate and old age components 
instead 
of only two (\cite[Torres-Papaqui et al. 2004]{Papaqui04}).

\newpage
\begin{acknowledgments}
We are pleased to thank the organizers for a very constructive, fruitful and
enjoyable Conference, and Lino Rodr\'\i guez, Emanuele Bertone, Sandro Bressan and
Miguel Ch\'avez, for providing their synthetic data prior to publication. 
\end{acknowledgments}

\end{document}